\documentclass[aps,prd,twocolumn,showpacs,nofootinbib]{revtex4}
\usepackage{latexsym}
\usepackage{amsmath,amsfonts}
\usepackage{amsbsy}
\usepackage{mathrsfs}
\usepackage{color}

\usepackage{psfrag}

\usepackage{enumerate}

\usepackage{amsmath,amssymb,calc,amsfonts}
\usepackage{latexsym}

\usepackage{graphicx,calc,epsfig}
\def\ut#1{\rlap{\lower1ex\hbox{$\sim$}}#1{}}
\newcommand{\N}{\mathbb{N}}

\newcommand{\be}{\nopagebreak[3]\begin{equation}}
\newcommand{\ee}{\end{equation}}
\newcommand{\ba}{\nopagebreak[3]\begin{eqnarray}}
\newcommand{\ea}{\end{eqnarray}}
\DeclareFontFamily{U}{rsfs}{}         
\DeclareFontShape{U}{rsfs}{m}{n}{<5> rsfs5 <6><7> rsfs7          %
  <8><9><10><10.95><12><14.4><17.28><20.74><24.88> rsfs10}{}     %
\DeclareMathAlphabet{\mathfs}{U}{rsfs}{m}{n}                     %
\newcommand{\n}{{\nonumber}}
\newcommand{\va}{\scriptscriptstyle}

\newcommand{\nn}{\sqrt{j(j+1)}}
\newcommand{\gm}{g_{\va M}}

\def\pb#1{\rlap{\lower1.5ex\hbox{$\longleftarrow$}}{#1}}
\def\dpb#1{\rlap{\lower1.5ex\hbox{$\Longleftarrow$}}{#1}}
\def\spb#1{\rlap{\lower1.5ex\hbox{$\leftarrow$}}{#1}}
\def\sdpb#1{\rlap{\lower1.5ex\hbox{$\Leftarrow$}}{#1}}

\definecolor{blue}{rgb}{0,0,1}
\definecolor{green}{rgb}{0,1,0}
\definecolor{red}{rgb}{1,0,0}
\definecolor{vio}{rgb}{1,0,1}
\definecolor{ama}{rgb}{1,1,0}

\begin{document}

%
%



\title{The scaling of black hole entropy in loop quantum gravity}

\date{\today}

\author{Amit Ghosh$^1$}
\author{Alejandro Perez$^2$}

\affiliation{$^1$Saha Institute of Nuclear Physics, 1/AF Bidhan Nagar,
700064 Kolkata, India.\\
$^2$Centre de Physique Th\'eorique, Campus de Luminy, 13288
Marseille, France.}


\vskip-3cm
\begin{abstract}
We discuss some general properties of black hole entropy in loop quantum gravity from the perspective of local stationary observers at distance $\ell$ from the horizon. The present status of the theory indicates that black hole entropy differs from the low energy (IR) expected value $A/(4G)$ (in natural units) in the deep Planckian regime (UV). The partition function is well defined if the number of non-geometric degrees of freedom $g_{\va M}$ (encoding the degeneracy of the area $a_p$ eigenvalue at a puncture $p$) satisfy the holographic bound $g_{\va M}< \exp(a_p/(4G))$. Our framework provides a natural renormalization mechanism such that 
$S_{UV}\to S_{IR}=A/(4 G_{Newton})$ as the scale $\ell$ flows.
\end{abstract}


\maketitle

\section{Introduction}

One of the main results of loop quantum gravity (LQG) \cite{lqg} is that the operators associated with area, volume and such geometric quantities have discrete spectra with a gap controlled by the length scale $\ell_g^2=\gamma G$  given by the  product of Newton's constant $G$ and the Immirzi parameter $\gamma$ \cite{Immirzi} (we use $\hbar=c=1$). This property of geometric operators is at the heart of the computation of black hole entropy using statistical mechanical methods in LQG studied in the past \cite{Corichi:2009wn}. In all these works an entropy proportional to the black hole horizon is found; however, the proportionality constant depends on the  details of the models used and is generically a function of the Immirzi parameter. The expected semiclassical Hawking black hole entropy $S=A/(4G)$ in such framework can only be recovered by the fine-tuning  of the Immirzi parameter.  

As the Hawking entropy of a black hole is a property of the low energy gravitational theory it has been argued in  \cite{jacob} that the renormalization of Newtons constant must be taken into account.  
In this work we adopt the general perspective of \cite{jacob}  and study the problem using the recently introduced local formulation of quantum horizons in LQG \cite{us}.
This new local perspective will allow us to argue in more detail how the semiclassical Hawking black hole entropy is expected to arise from the renormalization group flow 
towards low energies.

In order to describe our assumptions in more detail let us come back to the LQG prediction of discreteness of geometry at Planck scale. 
One of the striking properties of such discreteness is that it is independent of any matter content and any dynamics of quantum gravity: it is in fact a simple consequence of the canonical commutation relations which are directly derived from the symplectic structure associated with the geometric part of the classical gravity action. Thus there is a question as to what is the correct  value of the above couplings  that sets the scale of discreteness $\ell_g^2=\gamma G$ (is it the classical values, bare values or else?). We will see that this question is very important for the consistency of the quantum treatment of black holes in loop quantum gravity. The analysis presented in this paper makes it necessary to face this question head on. 
 
The general wisdom that comes from the study of interacting quantum field theories tells us that the couplings that are relevant for low energy phenomena are not necessarily the same as the ones that are involved in the high energy dynamics. In fact the coupling constants flow with energy scales via the renormalization group. From this point of view it seems unlikely that the fundamental discreteness discovered in LQG (at a scale $\ell_g$) can be described by the classical value of Newton's constant (relevant for solar system Physics). The view maintained in this work is that the fundamental scale of discreteness is set by the values of the relevant couplings at the UV cut-off for gravity (whose existence is necessary in the framework of LQG). As the notion of renormalization group flow is a dynamical issue (integrating out the dynamical degrees of freedom) a precise understanding of such UV regime is to be established from the study of the quantum constraints 
 in the dynamical description of LQG. 

Why is the dynamics relevant if one finds out discrete spectra of geometric operators directly from the classical symplectic structure? It is important because the geometric operators that have these features are merely kinematical observables---they do not commute for instance with the Hamiltonian constraint (encoding gauge invariance and dynamics in the background independent context). According to the standard Dirac canonical formulation these quantities are not observables due to their lack of gauge invariance. The true geometric observables must commute with the Hamiltonian constraint and in this way the nature of their spectra get intertwined with the dynamics.

It is usually expected that the discreteness of the kinematical observables will translate into the discreteness of geometric Dirac observables (for a discussion of this point see \cite{dirac}). Heuristically, and at the classical level, it is not difficult to imagine such Dirac observables. For instance, the area of a region defined by some function of the coordinates is clearly not a Dirac observable while the area of the surface of a black hole horizon is. Only the second has an invariant meaning while the first is just a coordinate dependent quantity. The quantization referred to above concerns the first example. Its validity in the second case (or the appropriate analogue in the QFT context, which in itself is a difficult open question) must depend on the interplay between geometry and matter degrees of freedom and hence is a fully dynamical question. 

There is at least one example where the question of the invariant definition of a geometric observable seems tractable at present---this is the area of the isolated horizon system describing a black hole horizon that is sufficiently large and isolated so that it may be well approximated by a stationary near horizon geometry. Here we will assume that the BH area is quantized according to LQG and that the relevant discreteness scale is fixed by the values of $\gamma$ and $G$ at the UV cut-off denoted by $\gamma_*$ and $G_*$ respectively. In other words, we assume that, in the case of the BH horizon area, the details of the quantum constraint dynamical requirements discussed above are fully encoded in the cut-off values of the relevant couplings. These assumptions combined with a recently introduced local formulation of quantum horizons in equilibrium \cite{ernesto, us} will lead to important insights into the nature of the IR flow of BH entropy in LQG and its consistency with semiclassical approaches. Here we are revisiting some aspects of this with the recently developed perspective. We believe that these results will shed new lights on the general question of the continuum limit in LQG.   

We believe that the analysis of the issues in black hole physics, where so much is known from both classical and semiclassical (QFT on curved spacetimes) analysis, is the ideal laboratory to start looking into the difficult questions concerning the quantum dynamics of LQG.

\section{Black hole entropy in LQG: consequences of the area hamiltonian}

Our system is a large isolated black hole. It is isolated in the precise sense that its classical near-horizon geometry is well approximated by an isolated horizon \cite{IH}. The area of the horizon is denoted by $A$. We describe the system in the reference frame of local observers who are stationary at a fixed proper distance $\ell$ from the horizon in the near-horizon geometry. The scale $\ell$ is assumed to be such that $G(\ell)\lesssim\ell^2\ll A$ where $G(\ell)$ is the value of the gravitational coupling which determines the Planck length at scale $\ell$.

As shown in \cite{ernesto}, in the reference frame of these local observers the quasilocal energy of the black hole is proportional to its area $A$. More precisely, the Hamiltonian is given by ($G$ is replaced by $G(\ell)$ in the Hamiltonian obtained in \cite{ernesto} and in units of $c=\hbar=1$) 
\be\label{he}
H(\ell)=\frac{A}{8\pi\ell G(\ell)}.
\ee
We will study the statistical mechanics of all degrees of freedom relevant for these observers in the framework of canonical ensemble. We further assume that the local temperature measured by these observers is given by the Unruh temperature $T_{\va U}=1/(2\pi\ell)$ (this is a simple consequence of the form of near-horizon geometry of IHs and the validity of QFT in curved spacetime in the chosen range of $\ell$). A calculation of this temperature in the context of discrete geometry of LQG has been recently proposed by Bianchi \cite{Bianchi:2012ui,Bianchi:2012vp}.

The reference frame of the local observers depends on $\ell$. However, in accordance with LQG we assume the existence of a UV cut-off value for the Newton constant $G_*$ where the scale $\ell_*\sim\sqrt G_*$ plays the role of a UV cut-off. The Immirzi parameter $\gamma$ is expected to flow with $\ell$ as well. Thereby, the area spectrum of the horizon is quantized in units of $\gamma_*G_*$ where $\gamma_*$ is the value of $\gamma(\ell)$ at the UV cut-off $\ell_*$. States of the horizon are given by a collection of punctures (which can be seen as the endpoint of edges of spin-networks coming from the bulk of the spacetime) each carrying a non-vanishing spin $j$. If states are denoted by $|j_1,j_2\cdots\rangle$ then using the LQG area spectrum \cite{lqg} and (\ref{he}), we get 
\be
\widehat H(\ell)|j_1,j_2\cdots\rangle=[\frac{\gamma_*G_*}{\ell G(\ell)}  \sum_{p} \sqrt{j_p (j_p+1)}]|j_1,j_2\cdots\rangle.
\ee

Notice that the quantum degrees of freedom that are relevant for these observers are both geometric and non-geometric in nature. By non-geometric here we refer to those that do not contribute to the area eigenvalue and hence to the Hamiltonian (\ref{he}). For simplicity we will call the non-geometric degrees of freedom as {\em matter} but one should keep in mind that some of these could be gravitational in nature (e.g. gravitons). 

Here we come to a crucial implication of the form of the effective Hamiltonian: one cannot neglect matter degrees of freedom 
in the statistical mechanical description of the system. This is due to  the fact that the local temperature at scale $\ell$ (or equivalently local acceleration $a=1/\ell$) is sufficiently high for local observers for them to see particles of all energies present in a thermal bath according to the usual Planckian distribution. As the non-geometric degrees of freedom do not contribute directly to the Hamiltonian they enter into the canonical partition function only through the degeneracy factor of the area eigenvalue. This means that they behave as if they are at infinite temperature ($\beta=0$) at which all matter excitations are equally likely. 

Even when this follows directly 
from the form of the Hamiltonian (\ref{he}), one can give an intuitive explanation as follows: from the geometric point of view these degrees of freedom need in deed to be at infinite temperature in order that they are at equilibrium with the geometric degrees of freedom which are at a temperature $T_U$ due to the gravitational red-shift. A more complete physical picture is provided in \cite{ernesto} leading to the form of the Hamiltonian (\ref{he}) where one shows how the matter energy is translated into area when absorbed by the black hole.

Finally, we assume that matter and other non-area related degrees of freedom can be independently assigned to individual punctures---this is a natural choice in the framework of LQG because only at the punctures the geometry is excited. 

The inclusion of matter degrees of freedom in the analysis will lead to a non trivial flow of the entropy with the scale $\ell$ set by  the observers. This flow will make the predictions of LQG compatible with Hawking's area law at large scales. It is interesting that matter (which in our local framework must unavoidably be considered) appears as the key ingredient allowing the deep Planckian description of LQG to be consistent with the low energy QFT semiclassical treatment.  

Let us now go back to the calculation of the thermodynamical properties of our quantum system.
With the above assumptions the canonical partition function is given by
\ba\label{zeta}
Z=\sum_{\{s_j\}}\prod_j\frac{N!}{s_j!}\,\left[(2j+1)g_{\va M}\right]^{s_j}e^{-\beta s_jE_j}
\ea
where $E_j=\gamma_*G_*\nn/[\ell G(\ell)]$ (according to (\ref{he}) and the area spectrum of LQG \cite{lqg}), and $g_{\va M}\in \N$ is the degeneracy factor associated with the non-geometric degrees of freedom. The quantity $g_{\va M}$ parametrizes the microscopic degrees of freedom (matter degrees of freedom, gravitons etc. not affecting the area) associated with a single puncture of the horizon. Therefore, it is natural to assume the most general form for this factor
\be g_{\va M}=g_{\va M}(j,\gamma,\cdots;\ell),\ee 
where the dots denote all possible coupling constants of the matter degrees of freedom. A precise form of $\gm$ necessarily depends on the dynamics: in the canonical quantum framework of LQG this means that it will have to be characterized by the resolution of the quantum Hamiltonian constraints in the vicinity of the horizon of the order of $\ell$. We will not attempt to explore this difficult question in the present work (for some attempt see \cite{Pranzetti:2012pd}),  but still we will see in what follows that some generic properties can be obtained without a detailed knowledge of the $\gm$.

Going back to our computation of the partition function, a simple calculation gives
\be\label{zz}
\log Z=N\log[\sum_j(2j+1)g_{\va M}e^{-\beta E_j}].
\ee
The average energy $\label{energy}\langle E\rangle=-\partial\ln Z/\partial\beta$
at the inverse temperature $\beta_{\va U}=2\pi \ell$ is a function of $N,\ell$ and $g_{\va M}$. 
Notice that unless (for large $j$)
\be
g_{\va M}< \exp{(\frac{a_j}{4 G(\ell)})},
\ee
where $a_j=8\pi \gamma_*G_*\sqrt{j(j+1)}$ is the LQG area eigenvalue at a puncture with spin $j$, the partition function is not well defined. Therefore, non gravitational degrees of freedom must be bounded by  the above holographic bound. This is an important implication of the present analysis that will be discussed further below.

At thermal equilibrium one can obtain a relation between the number of punctures to the area
\be
{N}= \frac{A}{4G(\ell) a(\gamma,\cdots;\ell) }.\label{NArelation}
\ee
where we have defined
\be
 a=\frac{2\pi\frac{\gamma_*G_*}{G(\ell)} \sum_j(2j+1)g_{\va M}\sqrt{j(j+1)} e^{-2\pi \frac{\gamma_*G_*}{G(\ell)} \sqrt{j(j+1)}}  }{\sum_j(2j+1)g_{\va M}e^{-2\pi\frac{\gamma_*G_*}{G(\ell)} \sqrt{j(j+1)}}}.
\ee
The quantity $a$ is the one quarter of the one puncture area expectation value at $\beta_{\va U}$ in Planck units $G(\ell)$ (the peculiar $1/4$ factor is only introduced in our definitions for convenience in order to simplify the final expression of the entropy below (eq. (\ref{esta}))).
For the entropy we get
\begin{align}
\label{entro}
S=-\beta^2\frac{\partial}{\partial\beta}(\frac{1}{\beta}\log Z)=
\log Z+\beta\frac{A}{8 \pi\ell}.
\end{align}
At $\beta_{\va U}=2\pi \ell$, we get
\be
S=\frac{A}{4G(\ell)}+\Sigma[\gamma,\cdots;\ell] N
\ee
where 
\be
\Sigma[\gamma,\cdots;\ell]=\log[\sum_j(2j+1)g_{\va M}e^{-2\pi\frac{\gamma_*G_*}{G(\ell)} \sqrt{j(j+1)}}].
\ee
At thermal equilibrium we can make use of the equation of state (\ref{NArelation}) to rewrite the entropy in the 
Sackur-Tetrode form
\be\label{esta}
S=\frac{A}{4G(\ell)}\left(1+\frac{\Sigma[\gamma,\cdots;\ell]}{a(\gamma,\cdots;\ell)}\right).\ee 
A minimal semiclassical consistency of this entropy (which does depend on $\ell$ in a non trivial way) comes from the
fact that  (for stationary black holes)
\be \label{pri}\delta M=\frac{\ell_p^2 \kappa}{2\pi} \delta S+\Omega \delta J+\Phi \delta Q+\mu \delta N,
\ee
where $S$ is given by (\ref{esta}) and $\mu=-\kappa \ell^2_p\Sigma/(2\pi)$ is the chemical potential \cite{us}.
This naive semiclassical consistency follows from the fact that the above first law is exactly equivalent (as a simple calculation shows) to the
usual geometric first law $\delta M=\frac{\kappa}{2\pi} \delta (A/4)+\Omega \delta J+\Phi \delta Q$
for all values of $\ell$ \footnote{In a recent paper \cite{Smolin:2012ys} it has been argued that the well known argument by Jacobson \cite{Jacobson:1995ab}
on the possible relationship between Einstein's equation and thermodynamics can be reproduced from LQG.  This study is preformed under the assumption that $\mu=0$. We would like to point out that the above 
consistency between the geometric and the thermodynamical first laws (stated for more general null surfaces) is at the heart of the result. Thus we claim that the derivation of \cite{Smolin:2012ys} remains valid 
for non vanishing chemical potential.}.

\section{Renormalization and scaling} 

In the previous section we have seen that the entropy of the BH flows with $\ell$ in a nontrivial fashion and does not agree with the semiclassical entropy for arbitrary values of the scale $\ell$. Nevertheless, the entropy is completely consistent with the first law of black hole mechanics due to the appearance of a nontrivial chemical potential related to punctures in the local LQG description of the first law. However, the question remains as to how (i.e. in which limit) the semiclassical Hawking entropy is to be recovered. We will see in this section that agreement with the semiclassical entropy is achieved at large values of $\ell$ (IR/continuum limit) provided some conditions (renormalization conditions) are met by the renormalization group flow of the relevant couplings in the framework.

As the non-geometric degrees of freedom do not contribute to the energy, their thermal distribution is independent of the temperature. Physically this is as if  they were at ``infinite temperature". For this reason it is appropriate to view them as maximally degenerate which might lead one to think that $\gm=\infty$ when no quantum gravity effects are present. Now the quantum gravity effects that could regularize this divergence are necessarily non-perturbative in origin (continuum field configurations have infinitely many degrees of freedom to all orders of perturbation theory). In the previous section we have seen that unless $\gm<\exp{a_j/(4G(\ell))}$, the partition function and other quantities in the analysis are not defined. We shall therefore assume that matter is maximally degenerate in the sense that it is close to saturating the previous bound. 

Before making this statement more precise recall that according to the popular paradigm of coupling constant unification $G(\ell)$ flows to smaller values than $G_*$ for larger $\ell$. Even if this is not the case, that is $G(\ell)$ grows with $\ell$ but much slower than $\ell^2$, the ratio $G(\ell)/\ell^2$ is a small number in the IR (of course, $\ell^2\ll A$). Therefore, the degeneracy of matter can be expressed as an expansion in $G(\ell)/\ell^2$ in the IR, namely 
\ba\label{13} \n &&
\gm(j,...,\gamma,\ell)=\exp\left(\frac{a_j}{4G(\ell)}\right)\times\\ && \Big[\gm^0(j,...,\gamma)+\gm^1(j,...,\gamma) \frac{G(\ell)}{\ell^2}+\cdots\Big],
\ea
where the real numbers $\gm^i(j,...,\gamma)$ represent perturbative corrections to the holographic factor $\exp\left({a_j}/{(4G(\ell))}\right)$. Recall that the latter factor representing the maximal degeneracy of the non-geometric degrees of freedoms has been introduced here for the requirement of finiteness of the partition function discussed above. We expect that it should be possible to obtain a derivation of this factor from first principles, e.g. from the relevant solutions of the quantum Hamiltonian constraint. The restriction on the degeneracy of matter degrees of freedom presented above is a complete analogue of the Bekenstein-Bousso bound leading to the widespread notion of holography in the quantum gravity literature \cite{Bekenstein:1974ax, Bousso:1999xy}. It is interesting that such a bound comes out naturally in the present treatment. It is also insightful that it only concerns the matter degrees of freedom, just as in the usual setting where similar bounds
  were presented. For that reason, we refer to it as the {\em micro-holography} bound.

The function $\Sigma$ becomes
\ba
\Sigma&=&\log(z_0+z_1 \frac{G(\ell)}{\ell^2}+\cdots )\n\\
&=&\log z_0+\frac{z_1}{z_0} \frac{G(\ell)}{\ell^2}+\cdots  , \ea 
where $z_i=\sum(2j+1)\gm^i(j,...,\gamma)$.  Similarly, for $a$ we get,
\ba
\frac{1}{a}
&=&\frac{G(\ell)}{\gamma_*G_*a_0}\left(1+\frac{G(\ell)}{\ell^2} \frac{b_0z_1-z_0b_1}{z_0 b_0}+\cdots\right),
\ea
where  
\ba \label{bibi} {b_i}={2\pi }\sum_j(2j+1)\gm^i\sqrt{j(j+1)},\ea 
and $a_{0}=b_0/z_0$ is the IR value of $a$, namely $a\to a_0$ for $\ell^2\gg G(\ell)$.
Then the entropy at large scale $\ell$ is given by
\begin{align}\label{ese} S&=\frac{A}{4G(\ell)}\Big[1+\frac{G(\ell)}{\gamma_*G_* a_0}\Big({\log z_0}\nonumber\\
&+\frac{G(\ell)}{\ell^2}[\log z_0\frac{b_0z_1-z_0b_1}{b_0z_0}+\frac{z_1}{z_0}]+\cdots\Big)\Big].\end{align}
Notice that the entropy receives two types of quantum corrections in the infrared limit---one, the first correction term in (\ref{ese}) is a pure LQG correction coming from the chemical potential of the punctures (for more discussion on this point see below) and two, the second and subsequent correction terms in (\ref{ese}) coming from the matter (they are the analog of loops corrections in standard QFT). 

Now we impose a renormalization condition 
\be \label{RC}\frac{4G(\ell_{\va IR})S}{A}=1\ee 
where by $\ell_{\va IR}$ we mean the large $\ell$ limit. This is the deep IR region in our analysis. 
There are now two possibilities for the validity of the previous condition.
First, $a_0\gamma_* G_*\gg G(\ell_{\va IR})$, and $\ell^2\gg G(\ell)$ (this can happen in more than one ways: one, if gravity in the infrared is much weaker than in the UV, namely $G_*\gg G(\ell_{\va IR})$ and $a_0\gamma_*\sim o(1)$; two $G_*\lesssim G(\ell_{\va IR})$ but $a_0\gamma_*\gg 1$), the renormalization condition Eq. (\ref{RC}) is automatically satisfied. By definition 
\be G(\ell_{\va IR})\simeq G_{\rm Newton}.\label{irlimit}\ee
Second, if the condition $a_0\gamma_*G_*\gg G(\ell_{\va IR})$ is not satisfied by the renormalization flow then the renormalization condition  (\ref{RC}) requires that $z_0=1$ or more explicitly
\be\label{here}
1=\sum_{j} (2j+1) g_{\va M}^0(j,\cdots).
\ee
The previous equation resembles in form similar conditions on the Immirzi parameter found in the previous literature (see \cite{DiazPolo:2011np, BarberoG.:2012ae} and references therein). 
However, the non-perturbative pre-factor makes clear that all things considered, i.e., using what is known for matter 
degrees of freedom the value of the Immirzi parameter does not play any special role. In fact the previous equation 
puts condition on the IR value of all couplings present in the functional $g_{\va M}^0[j,\cdots]$.

Eq. (\ref{ese}) gives the total entropy and its quantum corrections. In order to recover the Bekentein-Hawking entropy in the IR we only need the condition (\ref{irlimit}) to be satisfied. This is clearly compatible with the expected flow from perturbation theory. However, notice that the full entropy, that in the IR flows to the Bekentein-Hawking expression, is given by (\ref{esta}) in the deep UV regime---it comes with a correction to the Bekenstein-Hawking expression involving the (local) chemical potential associated to number of punctures $\bar\mu=-T_{\va U}\Sigma$. According to our analysis the chemical potential flows to zero in the IR as
\be
\bar\mu=-\frac{\log z_0}{2\pi\ell}-\frac{G(\ell)}{\ell^3}\frac{z_1}{2\pi}+\cdots.
\ee 
This is consistent with the point of view that the number of punctures $N$ is a relevant thermodynamic variable only in the deep Planckian regime. For intermediate scales $\ell$ where perturbation theory and standard QFT can be applied, the role of puncture as a thermodynamic variable becomes less and less important because its chemical potential is very close to zero. This is very much expected on physical grounds. It says that the fundamentally discrete LQG description in the UV (in terms of polymer-like excitations, punctures, and chemical potential) is replaced by a continuum field theoretical description where $\mu$ is very close to zero. Notice that the non-trivial scaling of $S$ in (\ref{ese}) is due to nongeometric degrees of freedom has been observed earlier in the study of entanglement entropy \cite{solod} where the scale $\ell$ plays the role of a UV cut-off for the nongeometric degrees of freedom. 

According to the paradigm of entanglement entropy (\ref{ese}) is interpreted as a quantum correction \cite{Callan:1994py} to the geometric Bekenstein-Hawking entropy. In fact, according to this picture (basically illustrated in the paper \cite{solod}) the geometric $S_{\rm geo}$ plus matter contributions (appearing as entanglement entropy) $S_{\rm ent}$ combine to give the black hole entropy $S$
\be
S=S_{\rm geo}+S_{\rm ent}=\frac{A}{4G_R},
\ee
where $G_{R}$ is called the renormalized gravitational constant. From (\ref{ese}), 
\ba && \frac{1}{G_R(\ell)}=\frac{1}{G(\ell)}\Big[1+\frac{G(\ell)}{\gamma_*G_* a_0}\Big({\log z_0}\nonumber\\
&&+\frac{G(\ell)}{\ell^2}[\log z_0\frac{b_0z_1-z_0b_1}{b_0z_0}+\frac{z_1}{z_0}]+\cdots\Big)\Big].\ea
Note that in our view $G_R$ is merely a different variable to express the renormalization group flow of $G(\ell)$, while  the true Newton's constant is $G(\ell)$, not $G_R(\ell)$. 

Coming back to our expression (\ref{ese}), we have seen that the details of the scaling of the BH entropy depend on the renormalization of the various couplings including in particular $G(\ell)$. Such renormalization should take place at length scales where matter loop effects are important. Furthermore, matter loops may produce other sub-leading corrections to (\ref{13}), such as $\log\ell$. These corrections give rise to logarithmic corrections $\log A$ to the entropy (\ref{ese}). In the context of entanglement entropy log-corrections have been obtained by several authors \cite{ghosh,solod,ashoke}. These log-corrections should be compared with the log-corrections to $g_M$---in fact, it will be an important benchmark test of the semi-classical limit of LQG. Notice that these log-corrections are in general very different in origin from the so-called LQG log-corrections to (\ref{ese}). So it is not surprising that these corrections do not agree \cite{ashoke}---there is no reason f
 or these to agree. However, it is crucial that the log-corrections to $g_M$ and those in the entanglement entropy are compatible with each other. We intend to carry out these calculations in future. LQG also suggests that the matter contribution to entanglement entropy coming from sum over all loops is finite!

\section{Conclusions}

In this paper we have addressed the important question of semiclassical consistency of the black hole entropy in loop quantum gravity on the basis of some key assumptions, some of which are justified only recently. Let us conclude by reviewing these assumptions and discussing their implications for future investigations.

Of course, our main assumption is the validity of loop quantum gravity as a description of the  quantum degrees of freedom of gravity at Planckian scales. This hypothesis goes hand in hand with the expectation of the existence of a UV cut-off for gravity and thereby, the existence of well defined parameters $G_*$ and $\gamma_*$ at high energies.

The next important assumption is about the right effective Hamiltonian needed to describe the equilibrium statistical mechanics of the quantum horizon as seen by local stationary observers---given by (\ref{he}). While there is a direct calculation showing that such a simple Hamiltonian is the appropriate one at the classical level \cite{ernesto}, it is not obvious that one can extrapolate its validity to the quantum level. In this respect one must say that this is clearly the most natural choice from the point of view of loop quantum gravity where the area operator is one of the simplest object in the theory. Moreover, in the quantum description of isolated horizons only area quantum numbers appear in the characterization of physical states: local area quanta (spin of spin network edges ending at the horizon) is the only physical observables needed to define a complete set of commuting observables of the physical Hilbert space of the horizon \cite{Engle:2010kt}. Therefore, under the basic underlying assumption that 
LQG makes sense, area (or a function thereof) seems as the natural choice of local Hamiltonian. This together with the classical evidence mentioned above motivates this choice as a notion of energy.

The consequences of these basic inputs are quite important. It is clear that the deep Planckian expression of the black hole entropy obtained from the LQG framework does not need to agree with the semiclassical value predicted by Hawking's semiclassical calculation: the presence of a chemical potential associated with number of punctures is an unavoidable feature of the polymer-like nature of quantum states. In previous models agreement could be enforced by tuning the Immirzi parameter to a particular value. In all these models the possible contributions of matter fields at the Planck scale have been neglected for simplicity. However, as we have argued this simplifying assumption is hard to justify because matter fields as well as gravity are expected to be equally important at such high energies. In this paper we have shown that both problems can in principle be solved at one stroke with the introduction of the local formulation and the appropriate Hamiltonian (\ref{he}).

The above assumptions imply that the number of local non-gravitational degrees of freedom must satisfy a micro holographic bound given by the exponential of the one quarter of the area quantum at punctures in Planck units. Like the Bekenstein-Bousso bound, this bound concerns only non-geometric or matter degrees of freedom. If the bound is violated then the partition function and entropy are divergent and the analysis becomes inconsistent.  

Matter contribution to the entropy can also be interpreted as the entanglement entropy. Interestingly, if microholography is satisfied then the matter contribution is such that in our approach the usual {\em species} problem is resolved: entropy does not scale with the number of fields and matter contributions can be consistent with Hawking's semiclassical entropy for any number of species.  

Consistency with the semiclassical Hawking entropy puts nontrivial requirements on the behaviour of the renormalization group flow of the gravitation coupling $G(\l)$ and the other matter couplings. If $a_0\gamma_*G_*\gg G_{\rm Newton}$ then the chemical potential (the genuine LQG fundamental discreteness contribution) correction to Hawking entropy is negligible and semiclassical agreement is guaranteed at low energies. This is an interesting condition because it can be satisfied in at two different ways due to the fact that from (\ref{bibi}) it follows that  $a_0\ge \pi \sqrt{3}$. The first scenario is that the thermal distribution encoded in (\ref{zz}) is such that low spins dominate, i.e. $a_0$ is order unity. Then one needs the ratio of scales $G_{\rm Newton}/(\gamma_*G_*)\ll 1$ which might be expected from the properties of the renormalization group flow. The second scenario is the one where $G_{\rm Newton}/(\gamma_*G_*)$ is order one but the thermal distribution is dominated by large spins so that $a_0\gg 1$. 
In this second case the matter contribution $g_m$ to the partition function is the key question. This question should be settled by studying the details of the matter geometry interaction controlled by the quantum Hamiltonian constraint\footnote{It is interesting to note that the relationship between large spin dominance and the semiclassical limit is at the heart of the recent results on the semiclassical properties of the covariant path integral approaches (or spin foams) \cite{sfsc}.}.
If on the other hand the previous conditions are not satisfied then there remains a third possibility involving nontrivial conditions on the matter couplings (this is explicitly expressed in equation (\ref{here})). 

All these requirement can be seen as a nontrivial guiding principle to study matter interactions in LQG provided by our semiclassical understanding of black hole physics. We propose to view such requirement as a renormalization condition and to use Hawking's entropy as the analogue of a gravitational ``measurement'' fixing such condition. We hope that some future work using this prespective will shed new light 
on the difficult subject of the semiclassical and continuum limit of loop quantum gravity.

\section{Acknowledgements}
We would like to thank M. Han  and S. Speziale for discussions and inputs.

\end{document}